\documentclass[aps,pra,showpacs,superscriptaddress,preprint]{revtex4}
\usepackage{amssymb}
\usepackage{amsmath}
\usepackage{graphicx}

\catcode`ð=\active
 \defð{\u{g}}
 \catcode`Ð=\active
 \defÐ{\u{G}}
 \catcode`Ý=\active
\defÝ{\. I}
 \catcode`ö=\active
\defö{\"{o}}
 \catcode`Ö=\active
 \defÖ{\"O}
 \catcode`ü=\active
 \defü{\"{u}}
 \catcode`Ü=\active
 \defÜ{\"{U}}
 \catcode`Þ=\active
\defÞ{\c{S}}
 \catcode`þ=\active
 \defþ{\c{s}}
 \catcode`ý=\active
 \defý{{\i}}
 \catcode`ç=\active
\defç{\d{c}}
 \catcode`Ç=\active
\defÇ{\d{C}}

\input{tcilatex}
\begin{document}

\title{A quantum pseudodot system with two-dimensional pseudoharmonic
potential using the Nikiforov-Uvarov method}
\author{Sameer M. Ikhdair}
\email[E-mail: ]{sikhdair@neu.edu.tr}
\affiliation{Physics Department, Near East University, 922022 Nicosia, North Cyprus,
Turkey}
\author{Majid Hamzavi}
\email[E-mail: ]{majid.hamzavi@gmail.com}
\affiliation{Department of Basic Sciences, Shahrood Branch, Islamic Azad University,
Shahrood, Iran}
\date{%
\today%
}

\begin{abstract}
Using the Nikiforov-Uvarov (NU) method, the energy levels and the wave
functions of an electron confined in a two-dimensional (2D) pseudoharmonic
quantum dot are calculated under the influence of temperature and an
external magnetic field inside dot and Aharonov-Bohm (AB) field inside a
pseudodot. The exact solutions for energy eigenvalues and wave functions are
computed as functions of the chemical potential parameters, applied magnetic
field strength, AB flux field, magnetic quantum number and temperature.
Analytical expression for the light interband absorption coefficient and
absorption threshold frequency are found as functions of applied magnetic
field and geometrical size of quantum pseudodot. The temperature dependence
energy levels for GaAs semiconductor are also calculated.

\end{abstract}

\maketitle

\newpage

\section{Introduction}

Over a long time, a considerable interest has been paid in studying size
effects in orbital magnetism [1,2] and the magnetic properties of
low-dimensional (2D) metallic and semiconducting structures with restricted
geometries [3] on nanostructures such as dots, wires, wells, antidots, well
wires and antiwells [4,5,6]. These structures can confine charge carriers in
one, two and three dimensions. Experimental research is currently made to
study the optical and quantum properties of low-dimensional semiconducting
structures for the fabrication purposes and subsequent working of electronic
and optical devices. More studies analyzing these structures have been
focused on the interband light absorption coefficient in the spherical
[7,8,9], parabolic, cylindrical and rectangular [10]\ quantum dots under the
influence of magnetic field [11,12]. More other works on optical properties
in nanostructures [13,14], band structure calculations, transport properties
of Aharonov-Bohm (AB) type oscillations [15,16] and
Altshuler-Aharonov-Spivak type oscillation [17,18].

The quantum antidot structure has been modeled using the repulsive antidot
harmonic oscillator with external magnetic and AB fields in cylindrical
coordinates to obtain an exact bound state solutions for the Schr\"{o}dinger
equation. In addition, the influence of dots and antidots on thermodynamic
properties (e.g., magnetization) of the system, the magnetic transport
properties and also the magneto-optical (MO) spectroscopic characteristics
of a 2D electron gas in a magnetic field are studied [19]. The nature of MO
transitions in this system demonstrate the appearance of rich spectrum of
nonequidistant frequencies are different from the MO spectrum for a dot
modeled by a harmonic potential. The quantum antidot is modeled as an
electron moving outside a cylinder of radius $a$ in the presence of magnetic
and AB flux fields to find analytical expressions for energy and wave
function [20]. The intensive investigations have shown that optoelectronic
properties of quantum dots are quite sensitive to the reduction of their
dimensionality and to the strength of applied external magnetic field, and
depend strongly on the electron-electron interaction. The numerical and
analytical solutions obtained for the dynamics of two classical electrons
interacting via a Coulomb field in a 2D antidot superlattice potential in
the presence of crossed electric and magnetic fields are quite different
than the noninteracting electrons [21]. Some authors have studied a 2D
theoretical model for the quantum dots in which electrons were confined by a
nonhomogenous magnetic field (the so-called magnetic antidot) [22]. The
pseudoharmonic (PH) interaction [23,24] is used in modeling the quantum dots
(QDs) and quantum antidots (QADs) in the presence of a strong magnetic field
together with an AB flux field in nanostructures [25]. The spectral
properties in a 2D electron confined by a pseudoharmonic quantum dots
(PHQDs) potential under the influence of a uniform magnetic field $%
\overrightarrow{B}$ along the $z$ direction and AB flux field created by a
solenoid inserted inside the pseudodot have been studied [25]. The
electron-phonon interaction on the surface of a sphere has been investigated
in the presence of a uniform strong magnetic field [9]. The energies of
ground and excited states have been analyzed with respect to electron-phonon
coupling constant$,$ sphere radius and dimensionless magnetic field [9].
Furthermore, the phonon interactions to the energy levels of a 2D electron
confined by a parabolic potential in the presence of an antidot potential
that produces a radially symmetrical hole have been studied in the presence
of a uniform magnetic field along the $z$ direction and AB flux created by a
solenoid inserted inside the antidot [26].

It is well-known that factors such as impurity, electric and magnetic
fields, pressure, and temperature play important roles in the electronic,
optical and transport properties of low-dimensional semiconductor
nanostructures [4,27-32]. Hence, many works in 2D quantum dots and
semiconductors are studied under the influence of external magnetic field
[33-38]. For example, the set of energy eigenstates of a 2D anisotropic
harmonic potential in a uniform magnetic field is found [33]. The formation
of dark states and the AB effect have been studied in
symmetrically/asymmetrically coupled three- and four-quantum dot systems.
Without a transverse magnetic field, destructive interference can trap an
electron in a dark state. However, the introduction of a transverse magnetic
field can disrupt the dark state giving rise to oscillation in current [34].
The propagator for an electron moving in 2D quadratic saddle-point potential
has been studied in the presence of a perpendicular uniform magnetic field
[35].\ The electron states in the 2D GaAs/AlGaAs quantum ring are
theoretically studied in effective mass approximation taking into account
on-centre donor impurity and uniform magnetic field perpendicular to the
ring plane [36]. Magneto transport properties of 2D electron gas in
AlGaN/AIN/GaN heterostructures have been studied [37]. The binding energy of
a hydrogenic impurity in self-assembled double quantum dots is calculated
via the finite-difference method. The variation in binding energy with donor
position, structure parameters and external magnetic field is studied [38].

Over the past years, the NU method [39] has shown to be a powerful tool in
solving the second-order differential equation. It was applied successfully
to a large number of potential models [40-43]. This method has also been
used to solve the Schr\"{o}dinger equation [40], relativistic spin-$0$ KG
equation [41], relativistic spin-$1/2$ Dirac equation [42] and spinless
Salpeter equation [43] with different potential models. Recently, an
alternative treatment is proposed for the calculations carried out within
the frame of the NU method which removes a drawback in the original theory
and bypasses some difficulties in solving the Schr\"{o}dinger equation [44].
This formalism has also been extended to the relativistic wave equations
[45]. The low-lying energy levels of two interacting electrons confined in a
2D parabolic quantum dot have been studied in magnetic field [45].

Therefore, we carry out detailed exact energetic spectrum and wave functions
of the Schr\"{o}dinger equation with a pseudoharmonic potential in the
presence of external magnetic and AB flux fields through the NU method [39].
The low-lying energy levels serve as a base for calculating the
corresponding interband light (optical) absorption coefficient and the
threshold frequency value of absorption for the given model. In addition,
the effect of the temperature on the effective mass is also calculated for
GaAs semiconductor.

The paper is organized as follows. In Sec. 2, we briefly present the basic
formulas of the NU method. In Sec. 3, the quantum dots and antidots with the
pseudoharmonic interaction are studied under the influence of external
magnetic and AB flux fields. The exact analytical expressions for the energy
spectrum and wave function are calculated. The interband light absorption
coefficient and temperature dependence of effective mass are also
investigated. Results and discussions are performed in Sec. 4. Finally, we
give our concluding remarks in Sec. 5.

\section{The Nikiforov-Uvarov Method}

The NU is usually used in solving a second-order hypergeometric-type
differential equations satisfying special orthogonal functions [39-43]. In
spherical or cylindrical coordinates, the resulting Schr\"{o}dinger-like
equation with a given potential is reduced to a hypergeometric type equation
through making a suitable change of variables, say, $r\rightarrow z$ and
then solved systematically for its exact or approximate eigensolutions
(energy levels and wave functions). The second-order hypergeometric equation
takes the form [39]%
\begin{equation}
\sigma ^{2}(z)g_{nl}^{\prime \prime }(z)+\sigma (z)\widetilde{\tau }%
(z)g_{nl}^{\prime }(z)+\widetilde{\sigma }(z)g_{nl}(z)=0,
\end{equation}%
where $\sigma (z)$ and $\widetilde{\sigma }(z)$ are at most second-degree
polynomials and $\widetilde{\tau }(s)$ is a first-degree polynomial. The
primes denote derivatives with respect to $z.$ To find a particular solution
of Eq. (1), one can use the wave functions, $g_{nl}(z)$ as 
\begin{equation}
g_{nl}(z)=\phi (z)y_{nl}(z),
\end{equation}%
to recast (1) into the hypergeometric-type equation 
\begin{equation}
\sigma (z)y_{nl}^{\prime \prime }(z)+\tau (z)y_{nl}^{\prime }(z)+\lambdabar
y_{nl}(z)=0,
\end{equation}%
where%
\begin{equation}
\lambdabar =k+\pi ^{\prime }(z),
\end{equation}%
and $y_{nl}(z)$ satisfies the Rodrigues relation:%
\begin{equation}
y_{nl}(z)=\frac{A_{n}}{\rho (z)}\frac{d^{n}}{dz^{n}}\left[ \sigma
^{n}(z)\rho (z)\right] .
\end{equation}%
In the above equation, $A_{n}$ is a constant related to the normalization
and $\rho (z)$ is the weight function satisfying the condition%
\begin{equation}
\left[ \sigma (z)\rho (z)\right] ^{\prime }=\tau (z)\rho (z),
\end{equation}%
with 
\begin{equation}
\tau (z)=\widetilde{\tau }(z)+2\pi (z),\tau ^{\prime }(z)<0.
\end{equation}%
Since $\rho (z)>0$ and $\sigma (z)>0,$ the derivative of $\tau (z)$ has to
be negative [39] which is the main essential condition in the choice of
particular solution relevant to the real bound state solution. The other
part of wave functions in Eq. (2) is the solution of the logarithmic
equation:%
\begin{equation}
\frac{\phi ^{\prime }(z)}{\phi (z)}=\frac{\pi (z)}{\sigma (z)},
\end{equation}%
where%
\begin{equation}
\pi (z)=\frac{1}{2}\left[ \sigma ^{\prime }(z)-\widetilde{\tau }(z)\right]
\pm \sqrt{\frac{1}{4}\left[ \sigma ^{\prime }(z)-\widetilde{\tau }(z)\right]
^{2}-\widetilde{\sigma }(z)+k\sigma (z)}.
\end{equation}%
is a polynomial of order one. The determination of $k$ is the essential
point in the calculation of $\pi (z),$ for which the discriminant of the
square root in the last equation is set to zero. This gives the polynomial $%
\pi (z)$ which is dependent on the transformation function $z(r).$ Also, the
parameter $\lambdabar $ defined in Eq. (4) takes the form%
\begin{equation}
\lambdabar =\lambdabar _{n}=-n\tau ^{\prime }(z)-\frac{1}{2}n\left(
n-1\right) \sigma ^{\prime \prime }(z),\ \ \ n=0,1,2,\cdots .
\end{equation}%
To obtain the energy formula, we need to establish a relationship between $%
\lambdabar $ and $\lambdabar _{n_{r}}$ by means of Eq. (4) and Eq. (10).

\section{ QDs and QADs in External Fields}

\subsection{Exactly solvable bound states}

Consider a 2D single charged electron, $e,$ with an effective mass, $\mu ,$
interacting via a radially symmetrical dot (electron) and antidot (hole)
potential in a uniform magnetic field, $\overrightarrow{B}=B\widehat{z}$ and
an AB flux field, applied simultanously. The Schr\"{o}dinger equation with
interaction potential field has the form [46]%
\begin{equation}
\left[ \frac{1}{2\mu }\left( \overrightarrow{p}+\frac{e}{c}\overrightarrow{A}%
\right) ^{2}+V_{\text{conf}}(\vec{r})\right] \psi (\vec{r},\phi )=E\psi (%
\vec{r},\phi ),
\end{equation}%
where $E$ is the energy eigenvalues, $\overrightarrow{p}=-i\hbar 
\overrightarrow{\nabla }$ is the momentum, $\mu $ is the effective mass of
an electron and $V_{\text{conf}}(\vec{r})$ is the scalar pseudoharmonic
interaction defined by [23,24] 
\begin{equation}
V_{\text{conf}}(\vec{r})=V_{0}\left( \frac{r}{r_{0}}-\frac{r_{0}}{r}\right)
^{2},
\end{equation}%
with $r_{0}$ and $V_{0}$ are the zero point (effective radius) and the
chemical potential. Besides, the vector potential $\overrightarrow{A}$ in
Eq. (11) may be represented as a sum of two terms, $\overrightarrow{A}=%
\overrightarrow{A}_{1}+\overrightarrow{A}_{2}$ having the azimuthal
components [25] 
\begin{subequations}
\begin{equation}
\overrightarrow{A}_{1}=\frac{Br}{2}\widehat{\phi },\text{ }\overrightarrow{A}%
_{2}=\frac{\Phi _{AB}}{2\pi r}\widehat{\phi },\text{ }\overrightarrow{A}%
=\left( \frac{Br}{2}+\frac{\Phi _{AB}}{2\pi r}\right) \widehat{\phi }.
\end{equation}%
\begin{equation}
\overrightarrow{\nabla }\times \overrightarrow{A}_{1}=\overrightarrow{B},%
\text{ }\overrightarrow{\nabla }\times \overrightarrow{A}_{2}=0,
\end{equation}%
where $\overrightarrow{B}$ $=B\widehat{z}$ is the applied magnetic field and 
$\overrightarrow{A}_{2}$ describes the additional magnetic flux $\Phi _{AB}$
created by a solenoid inserted inside the antidot (pseudodot). Let us take
the wave function $\psi (\vec{r},\phi )$ in cylindrical coordinates as 
\end{subequations}
\begin{equation}
\psi (\vec{r},\phi )=\frac{1}{\sqrt{2\pi }}e^{im\phi }g(r),\text{ }m=0,\pm
1,\pm 2,\ldots ,
\end{equation}%
where $m$ is the magnetic quantum number. Inserting the wave functions (14)
into the Schr\"{o}dinger equation (11), we obtain a second-order
differential equation satisfying $g(r),$%
\begin{equation}
g^{\prime \prime }(r)+\frac{1}{r}g^{\prime }(r)+\frac{1}{r^{2}}\left(
-\gamma ^{2}r^{4}+\nu ^{2}r^{2}-\beta ^{2}\right) g(r)=0,
\end{equation}%
with: 
\begin{subequations}
\begin{equation}
\nu ^{2}=\frac{2\mu }{\hbar ^{2}}\left( E+2V_{0}\right) -\frac{\mu \omega
_{c}}{\hbar }\left( m+\xi \right) ,
\end{equation}%
\begin{equation}
\beta ^{2}=\left( m+\xi \right) ^{2}+a^{2},
\end{equation}%
\begin{equation}
\gamma ^{2}=\frac{2\mu }{\hbar ^{2}}\frac{V_{0}}{r_{0}^{2}}+\left( \frac{\mu
\omega _{c}}{2\hbar }\right) ^{2},
\end{equation}%
where $\xi =\Phi _{AB}/\Phi _{0}$ is taken as integer with the flux quantum $%
\Phi _{0}=hc/e,$ $\omega _{c}=eB/\mu c$ is the cyclotron frequency and $%
a=k_{F}r_{0}$ with $k_{F}=\sqrt{2\mu V_{0}/\hbar ^{2}}$ is the fermi wave
vector of the electron. The magnetic quantum number $m$ relates to the
quantum number $\left\vert \beta \right\vert $ [Eq. (16b)].\tablenotemark[1]%
\tablenotetext[1]{For this system, only
two independent integer quantum numbers are required.} Further, the radial
wave function $g(r)$ has to satisfy the asymptotic behaviours,\textit{\ }%
that is,$\ $\ $g(0)\rightarrow 0$ and $g(\infty )\rightarrow 0.$ To make the
solution of Eq. (15) amendable by NU method, it is necessary to introduce
the following change of variables $s=r^{2},$ mapping $r\in (0,\infty )$ into
s$\in (0,\infty )$ which in turn recasts Eq. (15) into the hypergeometric
form (1) as 
\end{subequations}
\begin{equation}
g^{\prime \prime }(s)+\frac{2}{(2s)}g^{\prime }(s)+\frac{1}{(2s)^{2}}\left(
-\gamma ^{2}s^{2}+\nu ^{2}s-\beta ^{2}\right) g(s)=0.
\end{equation}%
Applying the basic ideas of Ref. [39], by comparing Eq. (17) with Eq. (1)
gives us the essential polynomials:%
\begin{equation}
\widetilde{\tau }(s)=2,~~~{\sigma }(s)=2s,~~~\widetilde{\sigma }(s)=-\gamma
^{2}s^{2}+\nu ^{2}s-\beta ^{2},
\end{equation}%
and substituting the polynomials given by Eq. (18) into Eq. (9), we obtain $%
\pi (s)$ as 
\begin{equation}
\pi (s)=\pm \frac{1}{2}\sqrt{\gamma ^{2}s^{2}+(2k-\nu ^{2})s+\beta ^{2}}.
\end{equation}%
The expression under the square root of the above equation must be the
square of a polynomial of first degree. This is possible only if its
discriminant is zero and the constant parameter (root) $k$ can be found by
the condition that the expression under the square root has a double zero.
Hence, $k$ is being obtained as $k_{+,-}=\nu ^{2}/2\pm \beta \gamma $. In
that case, it can be written in the four possible forms of $\pi (s)$; 
\begin{equation}
\pi (s)=\left\{ 
\begin{array}{cc}
+\left( \gamma s\pm \beta \right) , & \text{for }k_{+}=\frac{1}{2}\nu
^{2}+\beta \gamma , \\ 
-\left( \gamma s\pm \beta \right) , & \text{for }k_{-}=\frac{1}{2}\nu
^{2}-\beta \gamma .%
\end{array}%
\right.
\end{equation}%
One of the four possible forms of $\pi (s)$ must be chosen to obtain an
energy spectrum formula. Therefore, the most suitable physical choice is%
\begin{equation*}
\pi (s)=\beta -\gamma s,
\end{equation*}%
for $k_{-}$. The trick in this choice provides the negative derivative of $%
\tau (s)$ as required in Eq. (7). Hence, $\tau (s)$ and $\tau ^{\prime }(s)$
are obtained as 
\begin{equation}
\tau (s)=2\left( 1+\beta \right) -2\gamma s,\text{ }\tau ^{\prime
}(s)=-2\gamma <0~.
\end{equation}%
In this case, a new eigenvalue equation becomes 
\begin{equation}
\lambdabar _{n}=2\gamma n,\text{ }n=0,1,2,\ldots
\end{equation}%
where $\lambdabar _{n}=-n\tau ^{\prime }(s)-\frac{n(n-1)}{2}\sigma ^{\prime
\prime }(s)$ has been used and $n$ is the radial quantum number. Another
eigenvalue equation is obtained from the equality $\lambdabar =k+\pi
^{\prime }$ in Eq. (4), 
\begin{equation}
\lambdabar =\frac{\nu ^{2}}{2}-\gamma \left( \beta +1\right) .
\end{equation}%
In order to find an eigenvalue equation, the right-hand sides of Eq. (22)
and Eq. (23) must be compared with each other, i.e., $\lambdabar
_{n}=\lambdabar $. In this case the result obtained will depend on $E_{nm}$
in the closed form: 
\begin{equation}
\nu ^{2}=2\left( 2n+1+\beta \right) \gamma .
\end{equation}%
Upon the substitution of the terms of right-hand sides of Eqs. (16a)-(16c)
into Eq. (24), we can immediately arrive at the energy spectrum formula in
the presence of PH potential 
\begin{equation}
E_{nm}(\xi ,\beta )=\hbar \Omega \left( n+\frac{\left\vert \beta \right\vert
+1}{2}\right) +\frac{1}{2}\hbar \omega _{c}\left( m+\xi \right) -2V_{0},%
\text{ }\Omega =\sqrt{\omega _{c}^{2}+4\omega _{D}^{2}},
\end{equation}%
where $\left\vert \beta \right\vert $ is defined by Eq. (16b) and $\omega
_{D}=\sqrt{2V_{0}/\mu r_{0}^{2}}.$ We have two sets of quantum numbers $%
(n,m,\beta )$ and $(n^{\prime },m^{\prime },\beta ^{\prime })$ for dot
(electron) and antidot (hole), respectively. Therefore, energetic spectrum
formula (25) for the energy levels of the electron (hole) is identical to
Eq. (7) of Ref. [25] and usually used to study the thermodynamics properties
of quantum structures with dot and antidot in the presence and absence of
magnetic field.

We consider a few special cases of our results:

\begin{itemize}
\item Ignoring the last $-2V_{0}$ term, the above formula becomes the
Bogachek-Landman [19] energy levels in the presence of dot and antidot
potential.

\item In the absence of pseudoharmonic quantum dot (PHQD), i.e., $V_{0}=0,$ $%
\Omega \rightarrow \omega _{c},$ then $E_{nm}(\xi )=\hbar \omega _{c}\left[
n+\frac{1}{2}(\left\vert m+\xi \right\vert +1)\right] +\frac{1}{2}\hbar
\omega _{c}\left( m+\xi \right) $ which is the formula in the presence of $B$
and $\xi $ fields [19]$.$

\item When $\xi =0$ $(i.e.,\Phi _{AB}=0),$ we find the Landau energy levels,
i.e., $E_{nm}=\hbar \omega _{c}\left[ n+\frac{1}{2}(\left\vert m\right\vert
+m+1)\right] $.

\item When both $B=0$ ($\omega _{c}=0$) and $\xi =0$, we find $E_{nm}=\left(
4\hbar V_{0}/\mu r_{0}^{2}\right) \left[ n+\left( \sqrt{m^{2}+2\mu
V_{0}r_{0}^{2}/\hbar ^{2}}+1\right) /2\right] -2V_{0}.$

\item When $m=0$, we have $E_{n}=\left( 4\hbar V_{0}/\mu r_{0}^{2}\right)
\left( n+1/2\right) $ for harmonic oscillator energy spectrum.
\end{itemize}

Let us calculate the corresponding wave functions. We find the first part of
the wave function through Eq. (8), i.e.,%
\begin{equation}
\phi _{m}(s)=\exp \left( \int \frac{\pi (s)}{\sigma (s)}ds\right)
=s^{\left\vert \beta \right\vert /2}e^{-\gamma s/2}.
\end{equation}%
Then, the weight function defined by Eq. (6) as 
\begin{equation}
\rho (s)=\frac{1}{\sigma (s)}\exp \left( \int \frac{\tau (s)}{\sigma (s)}%
ds\right) =s^{\left\vert \beta \right\vert }e^{-\gamma s},
\end{equation}%
which gives the second part of the wave function (Rodrigues formula) via Eq.
(5), 
\begin{equation}
y_{n,m}(s)\sim s^{-\left\vert \beta \right\vert }e^{\gamma s}\frac{d^{n_{r}}%
}{ds^{n_{r}}}\left( s^{n+\left\vert \beta \right\vert }e^{-\gamma s}\right)
\sim L_{n}^{\left( \left\vert \beta \right\vert \right) }\left( \gamma
s\right) ,
\end{equation}%
where $L_{a}^{\left( b\right) }\left( x\right) =\frac{\left( a+b\right) !}{%
a!b!}F\left( a,b+1;x\right) $ is the associated Laguerre polynomial and $%
Fa,b;x)$ is the confluent hypergeometric function. The relation $g(s)=\phi
_{m}(s)y_{n,m}(s),$ gives the desired radial wave function as%
\begin{equation}
g(r)=C_{n,m}r^{\left\vert \beta \right\vert }e^{-\gamma r^{2}/2}F\left(
-n,\left\vert \beta \right\vert +1;\gamma r^{2}\right) ,
\end{equation}%
and hence the total wave function from Eq. (14) becomes%
\begin{equation*}
\psi _{n,m}(\vec{r},\phi )=\sqrt{\frac{\gamma ^{\left\vert \beta \right\vert
+1}n!}{\pi \left( n+\left\vert \beta \right\vert \right) !}}r^{\left\vert
\beta \right\vert }e^{-\gamma r^{2}/2}L_{n}^{\left( \left\vert \beta
\right\vert \right) }\left( \gamma r^{2}\right) e^{im\phi }
\end{equation*}%
\begin{equation}
=\frac{1}{\left\vert \beta \right\vert !}\sqrt{\frac{\gamma ^{\left\vert
\beta \right\vert +1}\left( n+\left\vert \beta \right\vert \right) !}{\pi n!}%
}r^{\left\vert \beta \right\vert }e^{-\gamma r^{2}/2}F\left( -n,\left\vert
\beta \right\vert +1;\gamma r^{2}\right) e^{im\phi }.
\end{equation}%
The energy levels in Eq. (25) differ from the usual Landau levels in
cylindrical coordinate system [47] to which it transforms when $\xi =0$, and 
$a\rightarrow 0$ (i.e., when the chemical potential of dot and antidot
vanishes, i.e., $V_{0}\rightarrow 0$). Nevertheless, the Landau levels are
nearly continuous discrete spectrum for a particle confined to a large box
with $B=0$ to equally spaced levels corresponding to $B>0.$ Each increment
of energy, $\hbar \omega _{c},$ corresponding to free particle states, which
is the degeneracy of each Landau level leading to a larger spacing as
magnetic field $B$ tends to become stronger [48]. The present model removes
this degeneracy with energy levels spectrum becomes%
\begin{equation}
E_{n,m}=\hbar \omega _{c}\left[ n+\frac{1}{2}\left( \left\vert m\right\vert
+m+1\right) \right] ,
\end{equation}%
and the wave function reads as%
\begin{equation}
\psi _{n,m}(\vec{r},\phi )=\frac{1}{m!}\sqrt{\frac{\gamma ^{m+1}\left(
n+m\right) !}{\pi n!}}r^{m}e^{-\gamma r^{2}/2}F\left( -n,m+1;\gamma
r^{2}\right) e^{im\phi },
\end{equation}%
where $\gamma =(\mu \omega _{c})/2\hbar .$ In the limit when $\omega _{c}\ll
g=\sqrt{\frac{8V_{0}}{\mu }}\frac{c}{r_{0}},$then we have%
\begin{equation}
E_{nm}=\varepsilon _{0}+\varepsilon _{1}\omega _{c}+\varepsilon _{2}\omega
_{c}^{2}-\varepsilon _{4}\omega _{c}^{4}+...,\text{ }
\end{equation}%
where%
\begin{equation}
\varepsilon _{0}=-2V_{0}+N_{nm}g,\text{ }\varepsilon _{1}=\frac{\hbar m}{2},%
\text{ }\varepsilon _{2}=\frac{N_{nm}}{2g},\text{ }\varepsilon _{4}=\frac{%
N_{nm}}{8g^{3}},\text{ }N_{nm}=\hbar \left( n+\frac{m+1}{2}\right) ,\text{ }%
g=\frac{1}{r_{0}}\sqrt{\frac{8V_{0}}{\mu }}.
\end{equation}

\subsection{Interband light absorption coefficient}

Expressions (25) and (30), obtained above for charge carriers (electron or
hole) energy formula and the corresponding wave function in quantum
pseudodot under the influence of external magnetic field and AB flux field,
allow to calculate the direct interband light absorption coefficient $K(%
\overline{\omega })$ in such system and the threshold frequency of
absorption. The light absorption coefficient can be expressed as [11-13,49]: 
\begin{equation*}
K(\overline{\omega })=N\dsum\limits_{n,m,\beta }\dsum\limits_{n^{\prime
},m^{\prime },\beta ^{\prime }}\left\vert \dint \psi _{n,m,\beta }^{e}(\vec{r%
},\phi )\psi _{n^{\prime },m^{\prime },\beta ^{\prime }}^{h}(\vec{r},\phi
)rdrd\phi \right\vert ^{2}\delta \left( \Delta -E_{n,m,\beta
}^{e}-E_{n^{\prime },m^{\prime },\beta ^{\prime }}^{h}\right) ,
\end{equation*}%
\begin{equation*}
=N\dsum\limits_{n,m,\beta }\dsum\limits_{n^{\prime },m^{\prime },\beta
^{\prime }}\frac{\gamma ^{\left\vert \beta \right\vert +\left\vert \beta
^{\prime }\right\vert +2}\left( n+\left\vert \beta \right\vert \right)
!\left( n^{\prime }+\left\vert \beta ^{\prime }\right\vert \right) !}{\pi
^{2}n!n^{\prime }!\left( \left\vert \beta \right\vert !\right) ^{2}\left(
\left\vert \beta ^{\prime }\right\vert !\right) ^{2}}\left\vert
\dint_{0}^{2\pi }e^{i\left( m+m^{\prime }\right) \phi }d\phi
\dint_{0}^{\infty }rdre^{-\left( \gamma +\gamma ^{\prime }\right)
r^{2}/2}r^{\left\vert \beta \right\vert +\left\vert \beta ^{\prime
}\right\vert }\right. \text{ }
\end{equation*}%
\begin{equation}
\times \left. F\left( -n,\left\vert \beta \right\vert +1;\gamma r^{2}\right)
F\left( -n^{\prime },\left\vert \beta ^{\prime }\right\vert +1;\gamma
^{\prime }r^{2}\right) \right\vert ^{2}\delta \left( \Delta -E_{n,m,\beta
}^{e}-E_{n^{\prime },m^{\prime },\beta ^{\prime }}^{h}\right) ,
\end{equation}%
where $\Delta =\hbar \overline{\omega }-\varepsilon _{g},$ $\varepsilon _{g}$
is the width of forbidden energy gap, $\overline{\omega }$ is the frequency
of incident light, $N$ is a quantity proportional to the square of dipole
moment matrix element modulus, $\psi ^{e(h)}$ is the wave function of the
electron (hole) and $E^{e(h)}$ is the corresponding energy of the electron
(hole).

Now, we proceed to calculate the light absorption coefficient [48-51]: 
\begin{equation}
K(\overline{\omega })=N\dsum\limits_{n,m,\beta }\dsum\limits_{n^{\prime
},m^{\prime },\beta ^{\prime }}P_{n,n^{\prime }}^{\beta }Q_{n,n^{\prime
}}^{\beta }\delta \left( \Delta -E_{n,m,\beta }^{e}-E_{n^{\prime },m^{\prime
},\beta ^{\prime }}^{h}\right) ,
\end{equation}%
where%
\begin{equation}
P_{n,n^{\prime }}^{\beta }=\frac{1}{\left( \left\vert \beta \right\vert
!\right) ^{4}}\left( \gamma \gamma ^{\prime }\right) ^{\left\vert \beta
\right\vert +1}\left( \frac{\gamma +\gamma ^{\prime }}{\gamma -\gamma
^{\prime }}\right) ^{2\left( n+n^{\prime }\right) }\frac{\left( n+\left\vert
\beta \right\vert \right) !\left( n^{\prime }+\left\vert \beta \right\vert
\right) !}{n!n^{\prime }!},
\end{equation}%
and%
\begin{equation}
Q_{n,n^{\prime }}^{\beta }=\left[ \left\vert \beta \right\vert !\left( \frac{%
2}{\gamma +\gamma ^{\prime }}\right) ^{\left\vert \beta \right\vert +1}%
\begin{array}{c}
_{2}F_{1}%
\end{array}%
\left( n,n^{\prime },\left\vert \beta \right\vert +1;-\frac{4\gamma \gamma
^{\prime }}{\left( \gamma -\gamma ^{\prime }\right) ^{2}}\right) \right]
^{2}.
\end{equation}%
Further, using Eqs. (25) and (35), we find the threshold frequency of
absorption as%
\begin{equation*}
\hbar \overline{\omega }=\varepsilon _{g}+\frac{\hbar }{2}\left( 2n+\sqrt{%
\left( m+\xi \right) ^{2}+2\mu V_{0}r_{0}^{2}/\hbar ^{2}}+1\right) \sqrt{%
\left( \frac{qB}{\mu c}\right) ^{2}+\frac{8V_{0}}{\mu r_{0}^{2}}}+\frac{%
q\hbar B}{2\mu c}\left( m+\xi \right)
\end{equation*}%
\begin{equation}
+\frac{\hbar }{2}\left( 2n^{\prime }+\sqrt{\left( m+\xi \right) ^{2}+2\mu
^{\prime }V_{0}r_{0}^{2}/\hbar ^{2}}+1\right) \sqrt{\left( \frac{qB}{\mu
^{\prime }c}\right) ^{2}+\frac{8V_{0}}{\mu ^{\prime }r_{0}^{2}}}+\frac{%
q\hbar B}{2\mu ^{\prime }c}\left( m+\xi \right) -4V_{0}.
\end{equation}%
where, $\xi =\Phi _{AB}/\Phi _{0}$ is an integer and $q=e.$ In the absence
of the AB flux field, i.e., when $\Phi _{AB}=0,$ we find that the threshold
frequency of absorption is identical to Eq. (22) of Ref. [49]. Further,
taking $n=m=0,$ then we have%
\begin{equation*}
\hbar \overline{\omega }_{00}=\varepsilon _{g}+\frac{\hbar }{2}\left( \sqrt{%
\xi ^{2}+2\mu V_{0}r_{0}^{2}/\hbar ^{2}}+1\right) \sqrt{\left( \frac{qB}{\mu
c}\right) ^{2}+\frac{8V_{0}}{\mu r_{0}^{2}}}+\frac{q\hbar B}{2\mu c}\xi
\end{equation*}%
\begin{equation}
+\frac{\hbar }{2}\left( \sqrt{\xi ^{2}+2\mu ^{\prime }V_{0}r_{0}^{2}/\hbar
^{2}}+1\right) \sqrt{\left( \frac{qB}{\mu ^{\prime }c}\right) ^{2}+\frac{%
8V_{0}}{\mu ^{\prime }r_{0}^{2}}}+\frac{q\hbar B}{2\mu ^{\prime }c}\xi
-4V_{0}.
\end{equation}%
in the presence of magnetic and AB fields.

For transition $000\rightarrow 000,$ the argument of Dirac delta function
allows one to define the threshold value of absorption as%
\begin{equation}
\frac{\hbar \overline{\omega }_{00}}{\varepsilon _{g}}=1+\frac{\left(
E_{0}^{e}+E_{0}^{h}\right) }{\varepsilon _{g}},
\end{equation}%
in which for quantum dot, we have 
\begin{subequations}
\begin{equation}
\frac{E_{0}^{e}}{\varepsilon _{g}}=\frac{1}{2}\left( \xi +1\right) \sqrt{%
k^{2}+\frac{8}{\rho ^{2}}}-\frac{2V_{0}}{\varepsilon _{g}},\text{ }
\end{equation}%
\begin{equation}
\frac{E_{0}^{h}}{\varepsilon _{g}}=\frac{1}{2}\left( \xi +1\right) \sqrt{%
k^{\prime 2}+\frac{8}{\rho ^{\prime 2}}}-\frac{2V_{0}}{\varepsilon _{g}},
\end{equation}%
\begin{equation}
\rho =\frac{r_{0}\varepsilon _{g}}{\hbar }\sqrt{\frac{\mu }{V_{0}}},\text{ }%
\rho ^{\prime }=\frac{r_{0}\varepsilon _{g}}{\hbar }\sqrt{\frac{\mu ^{\prime
}}{V_{0}}},\text{ }k=f(B)=\frac{e\hbar B}{\mu c\varepsilon _{g}},\text{ }%
k^{\prime }=f(B)=\frac{e\hbar B}{\mu ^{\prime }c\varepsilon _{g}}.
\end{equation}%
However, for quantum antidot,we have 
\end{subequations}
\begin{subequations}
\begin{equation}
\frac{E_{0}^{e}}{\varepsilon _{g}}=\frac{1}{2}\left( \sqrt{\xi ^{2}+2\mu
V_{0}r_{0}^{2}/\hbar ^{2}}+1\right) \left( \frac{e\hbar B}{\mu c\varepsilon
_{g}}\right) -\frac{2V_{0}}{\varepsilon _{g}},\text{ }
\end{equation}%
\begin{equation}
\frac{E_{0}^{h}}{\varepsilon _{g}}=\frac{1}{2}\left( \sqrt{\xi ^{2}+2\mu
^{\prime }V_{0}r_{0}^{2}/\hbar ^{2}}+1\right) \left( \frac{e\hbar B}{\mu
^{\prime }c\varepsilon _{g}}\right) -\frac{2V_{0}}{\varepsilon _{g}}.
\end{equation}%
In the absence of the AB flux field, i.e., $\xi =0,$ the above equations
(42a) and (42b) become identical to Eq. (27) of Ref. [49] and Ref. [12] in
quantum dot. Firstly, we study the variations of the threshold frequency of
absorption $\overline{\omega }_{00}$ (in units of $\varepsilon _{g}$) as a
function of magnetic field (in units of $k$). It is seen that $\overline{%
\omega }_{00}$ increases when the applied magnetic field increases (see
Figure 1). The effect of AB flux field on the interband energy is that the
lines remain linear but fan out or pushed up along the positive energy when $%
\xi $ increases. Secondly, the variations of the threshold frequency of
absorption $\overline{\omega }_{00}$ (in units of $\varepsilon _{g}$) with
quantum dot size (in units of $\rho $). It is seen in Figure 2 that $%
\overline{\omega }_{00}$ decreases when the quantum dot size increases.
However, it increases when the quantum pseudodot size increases [12,49].
Furthermore, the variations of $\overline{\omega }_{00}$ as a function of
magnetic field at small (large) applied $B$ is nonlinear (linear) as shown
in Figure 1 which is in agreement with Ref. [49] when $\xi =0$ . \ Finally,
in the presence of AB field, it changes linearly as $\xi $ increases.

\subsection{Temperature effect on effective mass and absorption threshold
frequency}

The variation of the effective mass with temperature is determined according
to the expression [32,52,53] 
\end{subequations}
\begin{equation}
\frac{\mu _{e}}{\mu (T)}=\frac{1}{f(T)}=1+E_{p}^{\Gamma }\left[ \frac{2}{%
E_{g}^{\Gamma }(T)}+\frac{1}{E_{g}^{\Gamma }(T)+\Delta _{0}}\right] ,
\end{equation}%
where $\mu _{e}$ is the electronic mass, $E_{p}^{\Gamma }=7.51$ $eV$ is the
energy related to the momentum matrix element, $\Delta _{0}=0.341$ $eV$ is
the spin-orbit splitting and $E_{g}^{\Gamma }(T)$ is the
temperature-dependence of the energy gap (in $eV$ units) at the $\Gamma $
point which is given by [13,52,54,55]%
\begin{equation}
E_{g}^{\Gamma }(T)=1.519-\frac{\left( 5.405\times 10^{-4}\right) T^{2}}{T+204%
}\text{ }(eV).
\end{equation}%
In Table 1, we display the temperature-dependent effective mass to the
effective mass of donor electron, i.e., $\mu (T)/\mu _{e}$ for different
values of temperatures. As seen in Table 1, the increase in the temperature
leading to a decrease in the value of $f(T)=\mu (T)/\mu _{e}.$ As a matter
of fact, the decrease in this value means that kinetic energy of the donor
electron decrease and consequently lowering the binding energy. The results
are similar to Ref. [32]. Hence the temperature dependence energy spectrum
formula can be expressed as%
\begin{equation}
E_{n,m}(B,T)=\frac{\hbar \omega _{c}}{f(T)}\left[ \sqrt{1+4\frac{\omega
_{D}^{2}}{\omega _{c}^{2}}f(T)}\left( n+\frac{\sqrt{\left( m+\xi \right)
^{2}+a^{2}f(T)}+1}{2}\right) +\frac{m+\xi }{2}\right] -2V_{0},
\end{equation}%
which for GaAs turns to be%
\begin{equation}
E_{n,m}(B,T)=14.9254\hbar \omega _{c}\left[ \sqrt{1+0.268\frac{\omega
_{D}^{2}}{\omega _{c}^{2}}}\left( n+\frac{\sqrt{\left( m+\xi \right)
^{2}+0.067a^{2}}+1}{2}\right) +\frac{m+\xi }{2}\right] -2V_{0},
\end{equation}%
where we have used $\mu =0.067\mu _{e}.$

To investigate the dependence of the energy levels on temperature, we take
the values of parameters: $B=6$ $T,$ $\xi =8$, $V_{0}=0.68346$ ($meV)$ and $%
r_{0}=8.958\times 10^{-6}$ $cm$ [25]. Hence, the temperature dependence of
the energy levels (in the units of $\hbar \omega _{c})$ at the $\Gamma $
point are given by 
\begin{equation*}
\frac{E_{n,m}(T)}{\hbar \omega _{c}}=\frac{1}{f(T)}\left[ \sqrt{1+\left(
0.32804\right) ^{2}f(T)}\left( n+\frac{\sqrt{\left( m+8\right) ^{2}+144f(T)}%
+1}{2}\right) +\frac{m+8}{2}\right]
\end{equation*}%
\begin{equation}
-1.9678584,
\end{equation}%
where $f(T)$ is calculated in Table 1 at any temperature value. In GaAs, we
have $f(T)=0.067$ [11]$.$ Taking the special values of parameters $\xi =8,$ $%
V_{0}=0.68459$ $meV$ and $r_{0}=8.958\times 10^{-6}$ $cm$ [25], two
parameters (temperature and magnetic field) dependence of the energy levels
(in units of $meV)$ are calculated as%
\begin{equation*}
E_{n,m}\left( B,T\right) =\frac{1}{f(T)}\left[ 0.1157705\sqrt{%
B^{2}+3.8803305f(T)}\left( n+\frac{\sqrt{\left( m+\xi \right) ^{2}+144f(T)}+1%
}{2}\right) \right.
\end{equation*}%
\begin{equation}
+\left. 0.1157705B\left( \frac{m+\xi }{2}\right) \right] -1.36918\text{ (}meV%
\text{),}
\end{equation}%
which becomes%
\begin{equation*}
E_{n,m}(B)=14.9254\left[ 0.1157705\sqrt{B^{2}+0.26}\left( n+\frac{\sqrt{%
\left( m+\xi \right) ^{2}+9.648}+1}{2}\right) \right.
\end{equation*}%
\begin{equation}
+\left. 0.1157705B\left( \frac{m+\xi }{2}\right) \right] -1.36918\text{ (}meV%
\text{)}.
\end{equation}%
When $n=0$ and $m=0,$ we obtain%
\begin{equation*}
E_{00}\left( B,T\right) =\frac{1}{2f(T)}\left[ 0.1157705\left( \sqrt{\xi
^{2}+144f(T)}+1\right) \sqrt{B^{2}+3.8803305f(T)}+0.1157705B\xi \right]
\end{equation*}%
\begin{equation}
-1.36918\text{ (}meV\text{),}
\end{equation}%
It is seen in Figure 3 that the energy of the ground state is linearly
changing with temperature for high temperatures when external fields are
abscent, i.e., $\xi =0$ and $B=0.$ As seen from Figure 4, for specific value
of temperature, the variation of the ground energy with the magnetic field
is linear (nonlinear) for large (small) magnetic field values. Furthermore,
the variation of the energy with the temperature increases with the
increasing temperature as demonstrated in Table 1 and Figure 4. For further
details see the Figures 1-5 given in Ref. [4] when we set $\xi =0$ in the
above energy formula.$.$

\section{Results and Discussions}

We solved exactly the Schr\"{o}dinger equation for an electron under the
pseudoharmonic interaction consisting of quantum dot potential and antidot
potential in the presence of a uniform strong magnetic field $%
\overrightarrow{B}$ along the $z$ axis and AB flux field created by an
infinitely long solenoid inserted inside the pseudodot. We have obtained
bound state solutions including the energy spectrum formula (25) and wave
function (30) for a Schr\"{o}dinger electron. Now we study the effect of the
pseudoharmonic potential, the presence and absence of magnetic field $B,$
the AB flux density $\xi $ and the antidot potential on the energy levels
(25). To see the dependence of the energy spectrum on the magnetic quantum
number, $m$, we take the following values: magnetic field $\overrightarrow{B}%
=\left( 6\text{ }T\right) $ $\widehat{z},$ AB flux field $\xi =8$, chemical
potential $V_{0}=0.68346$ ($meV)$ and $r_{0}=8.958\times 10^{-6}$ $cm$ [25].
Thus, we obtained $a=\sqrt{2\mu V_{0}r_{0}^{2}/\hbar ^{2}}=11.997702,$ $%
2\omega _{D}=\sqrt{8V_{0}/\mu r_{0}^{2}}=0.3280381$ $\omega _{c}$ and $\hbar
\omega =1.05243\hbar \omega _{c}$ with $\hbar \omega _{c}=0.1157705B$ (meV)
where $B$ is in Tesla [50]$,$ the dependence of the energy on the quantum
numbers $n$ and $m$ is given by%
\begin{equation}
\frac{E_{n,m}}{\hbar \omega _{c}}=1.05243\left( n+\frac{\sqrt{\left(
m+8\right) ^{2}+12^{2}}+1}{2}\right) +\frac{1}{2}\left( m+8\right)
-1.9678584,\text{ for }B=6\text{ }T.
\end{equation}%
where $m=0,\pm 1,\pm 2,\ldots $ and $n=0,1,2,\ldots .$ For the lowest ground
state ($n=0$): $E_{0,m}/\hbar \omega _{c}=1.05243\left( \sqrt{\left(
m+8\right) ^{2}+12^{2}}+1\right) /2+\left( m+8\right) /2-1.9678584,$ for $%
B=6 $ $T.$ Overmore, to show the effect of magnetic field $B$ on the energy
spectrum, we take values for parameters $\xi =8,$ $V_{0}=0.68459$ $meV$ and $%
r_{0}=8.958\times 10^{-6}$ $cm$ [25], where $a=\sqrt{2\mu
V_{0}r_{0}^{2}/\hbar ^{2}}=12.007617$ and $4\omega _{D}^{2}=8V_{0}/\mu
r_{0}^{2}=0.120039\times 10^{24}$ $(rad/s)^{2},$ the dependence of energy
levels on the magnetic field becomes%
\begin{equation*}
E_{n,m}\left( meV\right) =0.1157705\sqrt{B^{2}+3.8803305}\left( n+\frac{%
\sqrt{\left( m+\xi \right) ^{2}+12^{2}}+1}{2}\right)
\end{equation*}%
\begin{equation}
+0.1157705B\left( \frac{m+\xi }{2}\right) -1.36918,\text{ }\xi =8.
\end{equation}%
Note that in the abscence of AB flus field $\left( \xi =0\right) ,$ Eq. (52)
resembles Eq. (13) of Ref. [49].

In Figure 5, we plot the pseudodot energy levels in the absence (presence)
of pseudodot potential (i.e., $V_{0}=0$ $\rightarrow $ $a=0$ $\left(
V_{0}\neq 0\rightarrow a=12\right) )$ and in the absence (presence) of AB
flux field $\Phi _{AB}$ (i.e., $\xi =0$ ($\xi =8)$) as a function of
magnetic quantum number $m$ for $B=6$ $T$. As demonstrated in Figure 5, the
Landau energy states [47] (i.e., $V_{0}=0$ $\rightarrow $ $a=0,$ $\xi =0$
and $\ \xi =8)$ are degenerate states (see, long dashed and dotted solid
curves) for negative values of $m,$ however, the pseudodot potential removes
this degeneracy $($case when $V_{0}\neq 0$ $\rightarrow $ $a=12),$ (see,
solid and dotted dashed curves). In the absence of pseudodot potential ($a=0)
$ and presence of AB flux field ($\xi =8$), the degeneracy still exists
(long dashed line). It is found that the energy levels of PHQD potential are
approximately equal the Landau energy levels for large absolute $m$ values.
However, they are quite different for small absolute $m$ values ($-12\leq
m\leq 13$ when $\xi =0$ and $-20\leq m\leq 5$ when $\xi =8$)$.$ It is also
noted that as the quantum number $n$ increases ($n>0),$ the curves are quite
similar to Figure 5 but the energy levels are pushed up toward the positive
energy for all values of $m.$ In Figure 6a to Figure 6f, we plot the
magnetic field dependence of the ground state energy $E_{0,m}(\xi ,a)$ (in
units of $meV)$ in the presence and absence of pseudodot potential and AB
flux field for several values of magnetic quantum numbers $m=27,35,1,0,-24$
and $-16$, respectively. It is shown in Figure 6a to Figure 6f that
pseudodot energy increases with increasing magnetic field strength.
Furthermore, in the absence of pseudodot potential, magnetic field in the
positive $z$ direction removes the degeneracy for positive $m$ values. In
these Figures, the behavior of pseudodot energy as function of the magnetic
field $B$ is shown in the presence of pseudodot potential and AB flux field
(solid curves), in the absence of pseudodot potential and presence of AB
flux field (dotted curves) and the absence of pseudodot potential and AB
flux field (dashed curves).

for GaAs. In Figure 7, we show the variation of the pseudodot energy levels
(in $meV$) as function of magnetic field $B$ (in $Tesla$) (52). We consider
the cases: the presence of both pseudodot potential and $\xi $ (solid
curves), the absence of pseudodot potential and presence of $\xi $ (dotted
curves) and the absence of both pseudodot potential and $\xi $ (dashed
curves) taking the various values of radial $n$ and magnetic $m$ quantum
numbers. For GaAs case, we consider the following cases (a) $n=m=0,$ (b) $%
n=5,$ $m=0,$ (c) $n=0,$ $m=5,$ (d) $n=0,$ $m=-5$ and (e) $n=5,$ $m=-5$ in
Figure 7a to Figure 7e$,$ respectively$.$ In Figure 8, we plot the energy
levels (52) for various values of $n$ and $m$ quantum numbers as functions
of the magnetic field strength $B$ for the case $\xi =0$. It is seen that
the energy curves coincide with those obtained by Eq. (13) of Ref. [49]. We
also plot the case where $\xi \neq 0.$

\section{Concluding Remarks}

In this work, we have obtained bound state energy levels and wave functions
of the Schr\"{o}dinger particle in the 2D pseudoharmonic quantum dot and
antidot structure under the influence of external uniform magnetic and AB
flux fields. Overmore, the Schr\"{o}dinger bound state solutions are
obtained, in closed form, in the framework of the NU method. In our
application, we have calculated the energy and wave function solutions for a
few electrons bound at GaAs semiconductor interfaces whose velocities are no
relativistic. Overmore, the nonrelativistic electron and hole energy spectra
and the corresponding wave functions are used to calculate the interband
light absorption coefficient and the threshold frequency of absorption. This
energy spectrum of the electron (hole) may be also used in studying the
thermodynamics properties of quantum structures with dot (antidot) for
specific values of external uniform magnetic and AB flux fields and spatial
confinement length. The temperature dependence of the energy levels are
calculated in Table 1 at \ any temperature $T$ $($Kelvin).

\acknowledgments We thank the referees for their invaluable suggestions
which have greatly helped in improving the manuscript.

{\normalsize 
}

\bigskip

\baselineskip= 2\baselineskip
\bigskip \newpage

\bigskip

{\normalsize 
}

\baselineskip= 2\baselineskip

\begin{table}[tbp]
\caption{Calculated $f(T)$ with different values of temperature for GaAs. }%
\begin{tabular}{llll}
\tableline\tableline$T$ $(K)$ & $\mu (T)/\mu _{e}$ & $T$ $(K)$ & $\mu
(T)/\mu _{e}$ \\ 
\tableline$0$ & $0.0669984$ & $170$ & $0.0653679$ \\ 
$10$ & $0.0669886$ & $180$ & $0.0652177$ \\ 
$20$ & $0.0669608$ & $190$ & $0.0650643$ \\ 
$30$ & $0.0669174$ & $200$ & $0.0649080$ \\ 
$40$ & $0.0668603$ & $210$ & $0.0647490$ \\ 
$50$ & $0.0667911$ & $220$ & $0.0645874$ \\ 
$60$ & $0.0667112$ & $230$ & $0.0644235$ \\ 
$70$ & $0.0666217$ & $240$ & $0.0642573$ \\ 
$80$ & $0.0665236$ & $250$ & $0.0640891$\tablenotemark[1] \\ 
$90$ & $0.0664178$ & $260$ & $0.0639188$ \\ 
$100$ & $0.0663051$ & $270$ & $0.0637468$ \\ 
$110$ & $0.0661861$ & $280$ & $0.0635730$ \\ 
$120$ & $0.0660614$ & $290$ & $0.0633976$ \\ 
$130$ & $0.0659315$ & $300$ & $0.0632206$\tablenotemark[1] \\ 
$140$ & $0.0657968$ & $350$ & $0.0623154$ \\ 
$150$ & $0.0656577$\tablenotemark[1]\tablenotetext[1]{See Ref. [32].} & $400$
& $0.0613818$\tablenotemark[1] \\ 
$160$ & $0.0655147$ & $500$ & $0.0594513$\tablenotemark[1] \\ 
\tableline &  &  & 
\end{tabular}%
\end{table}

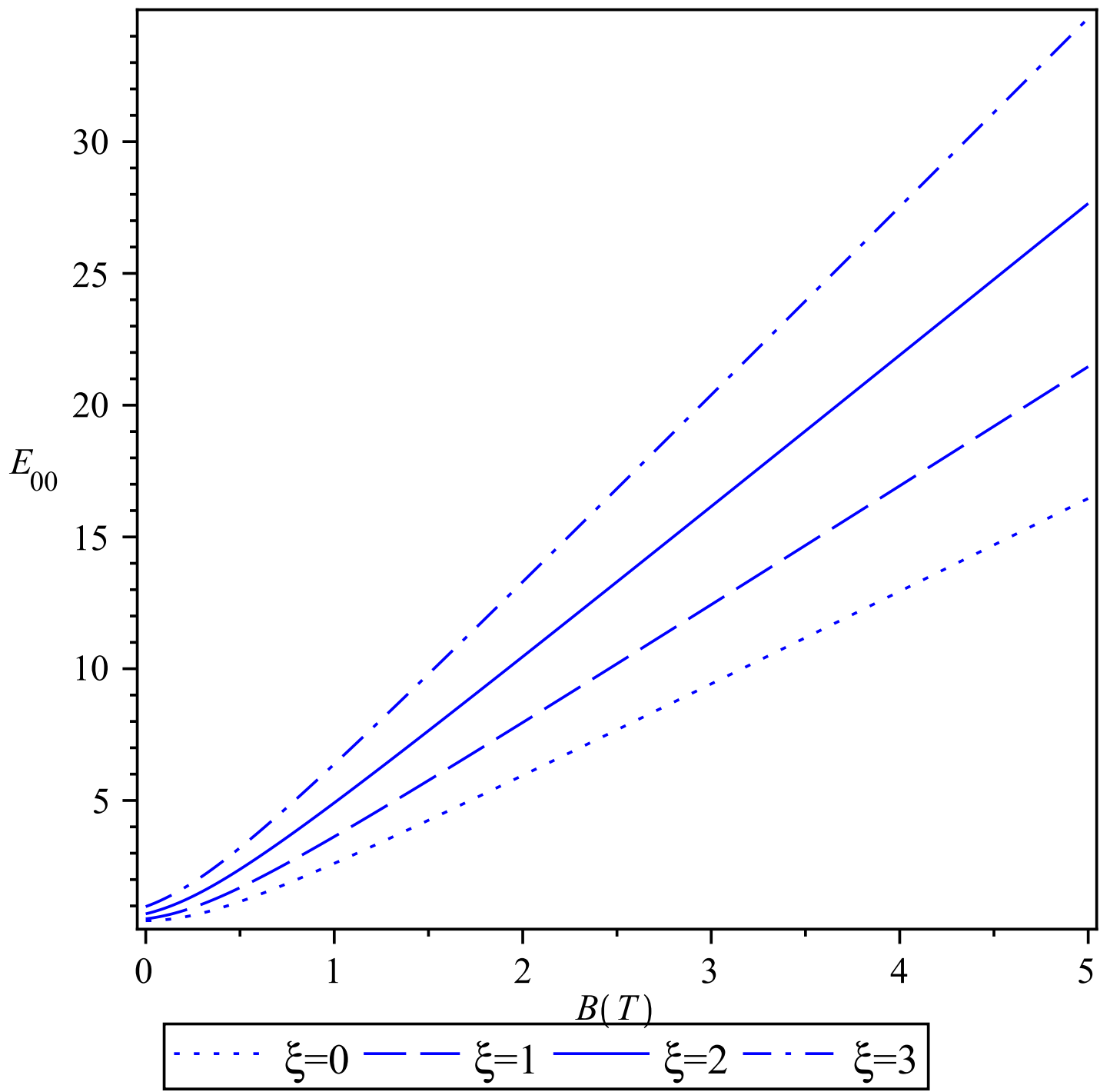
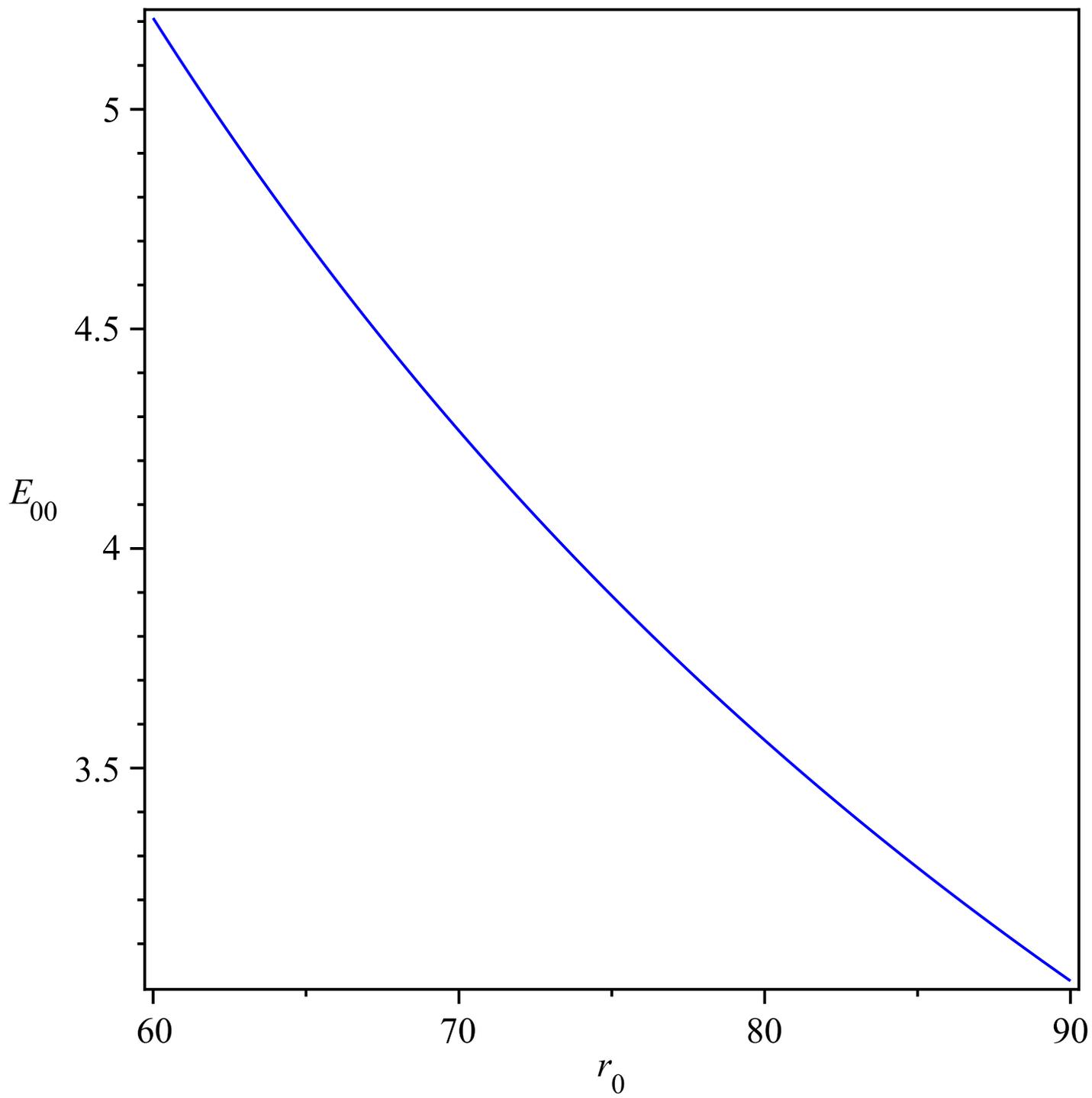
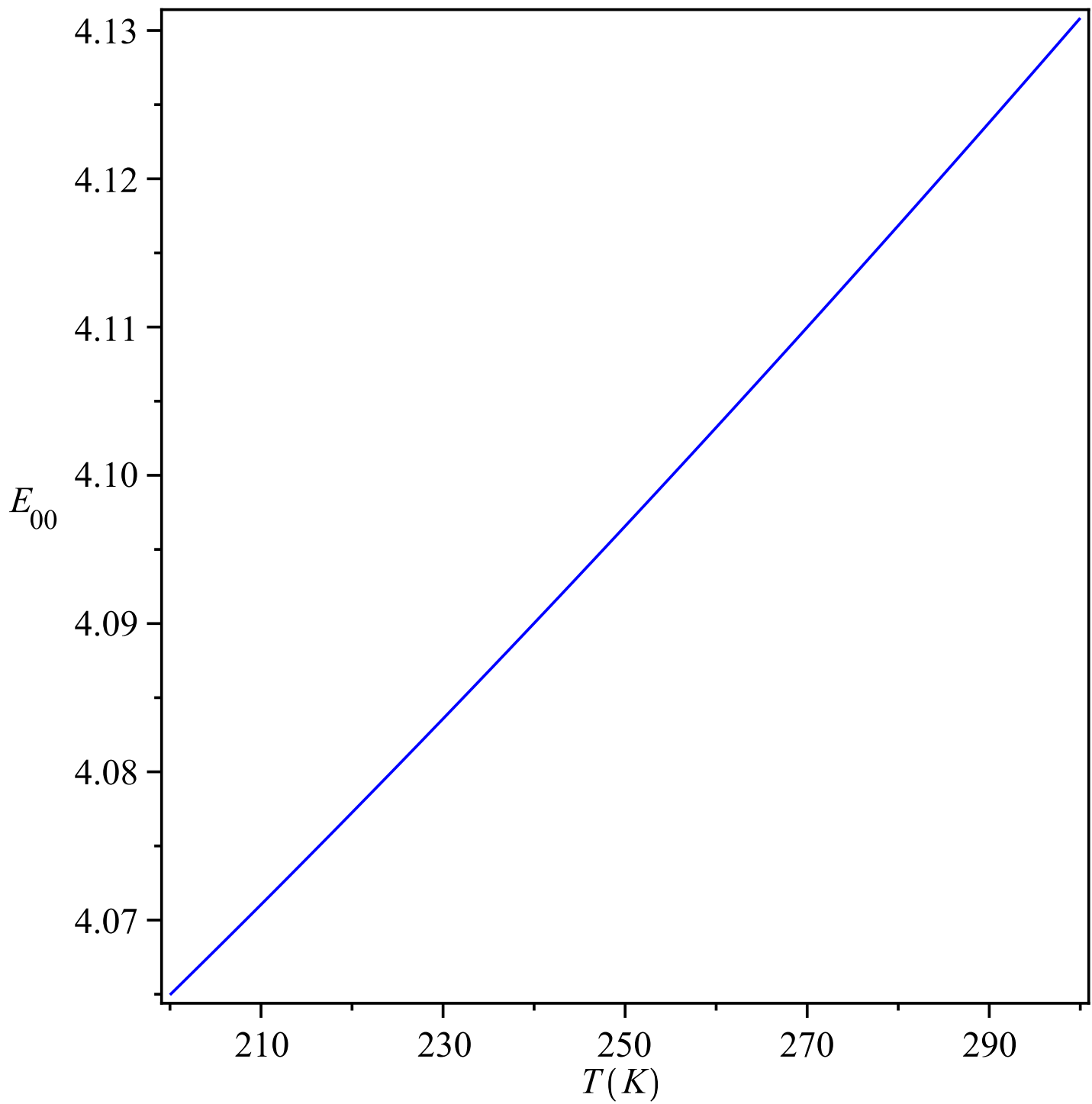
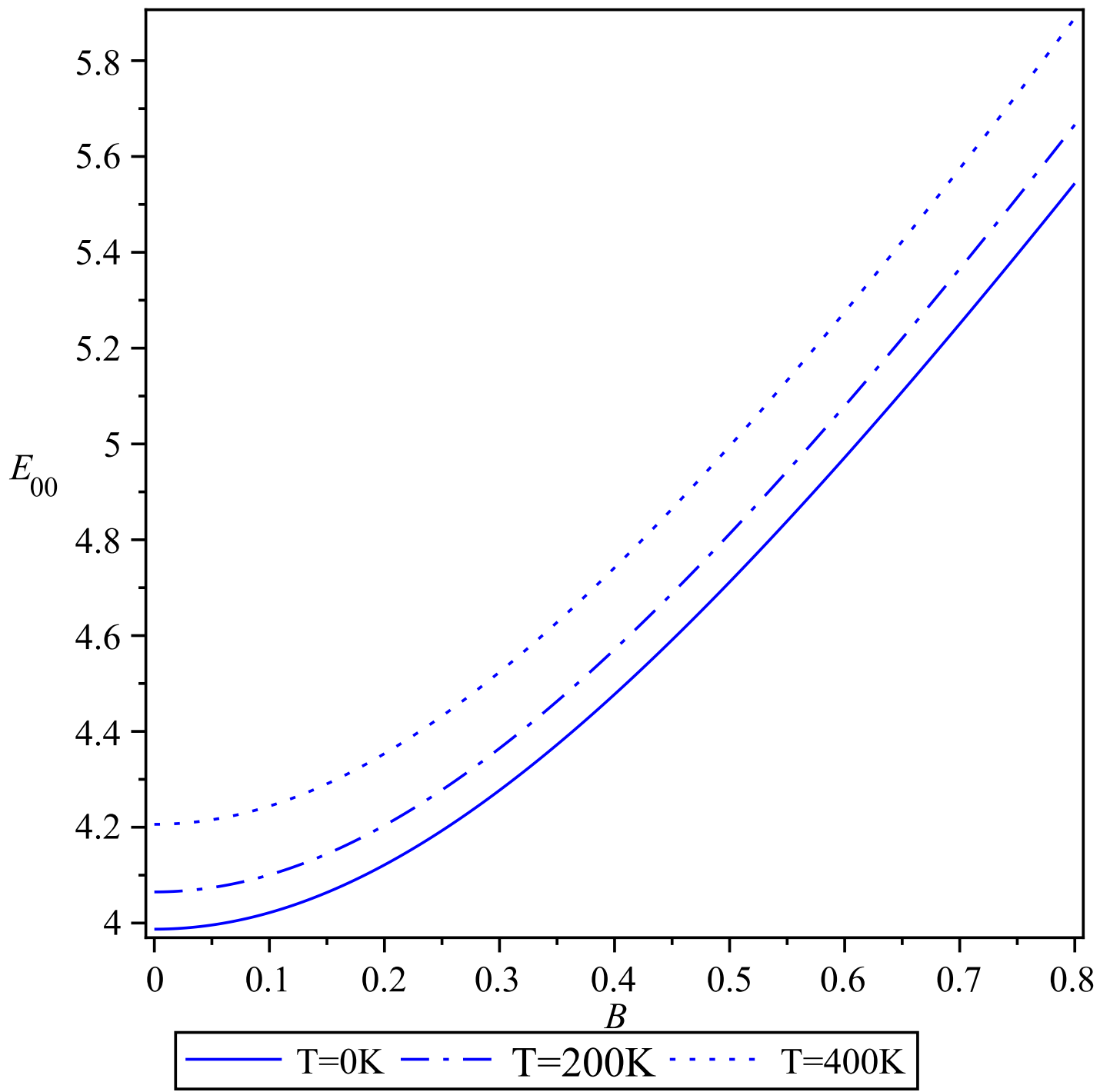
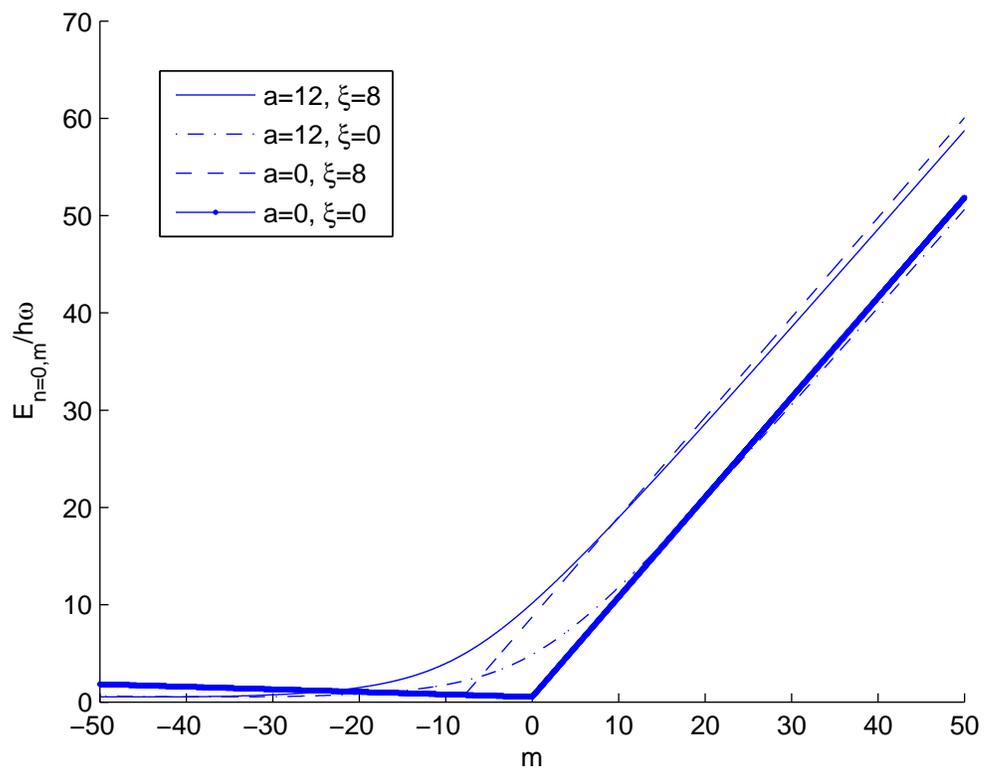
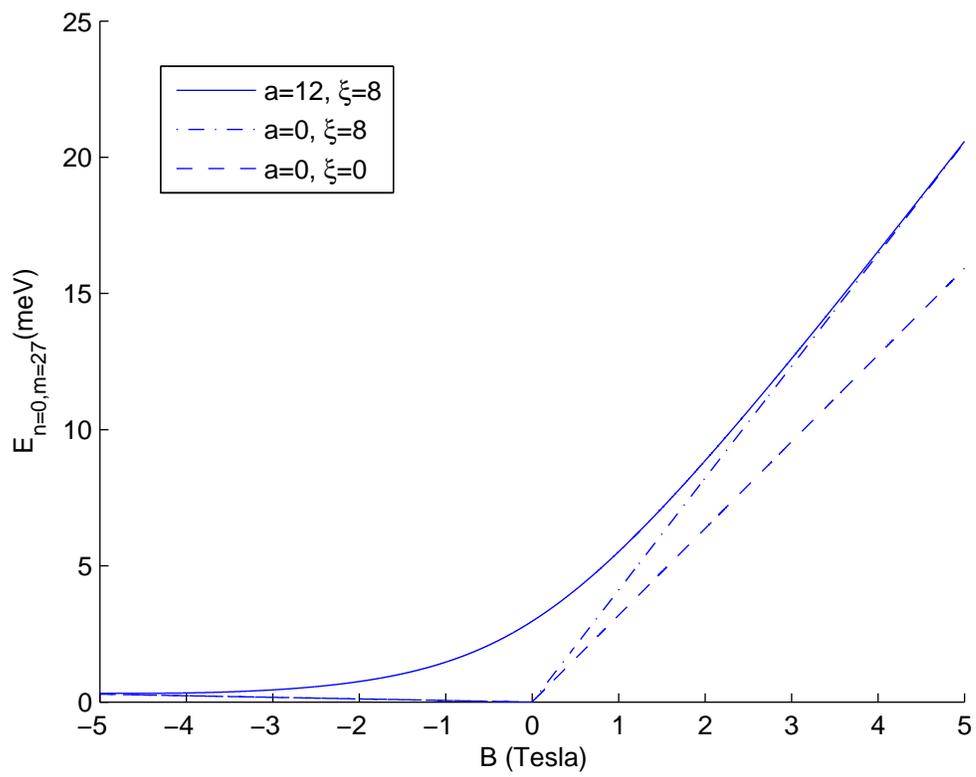
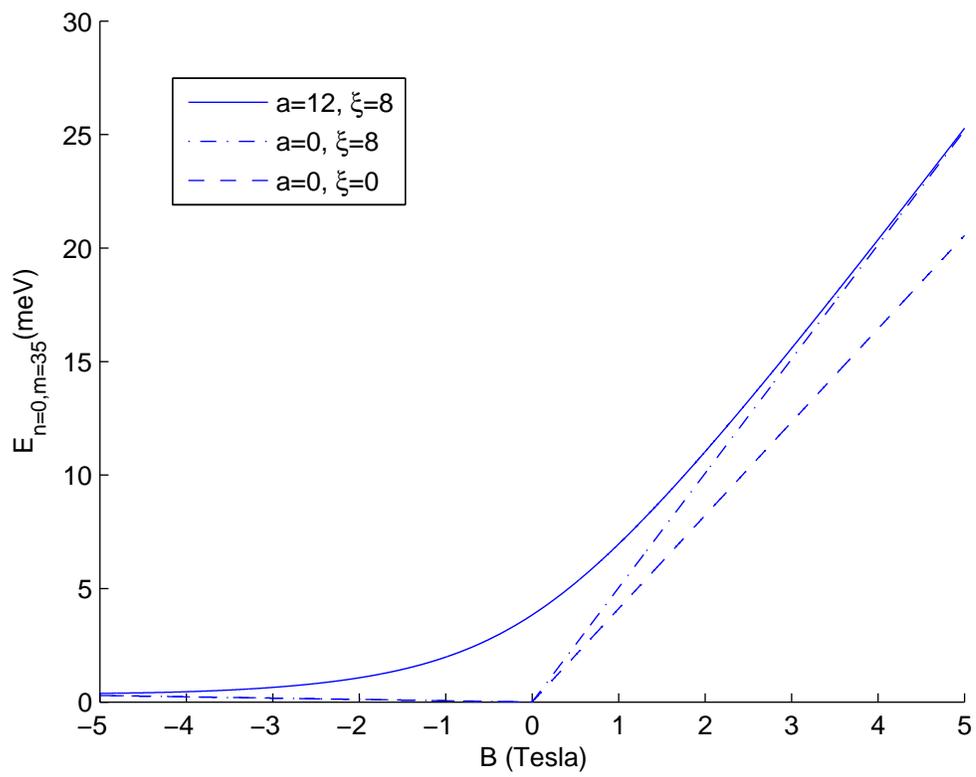
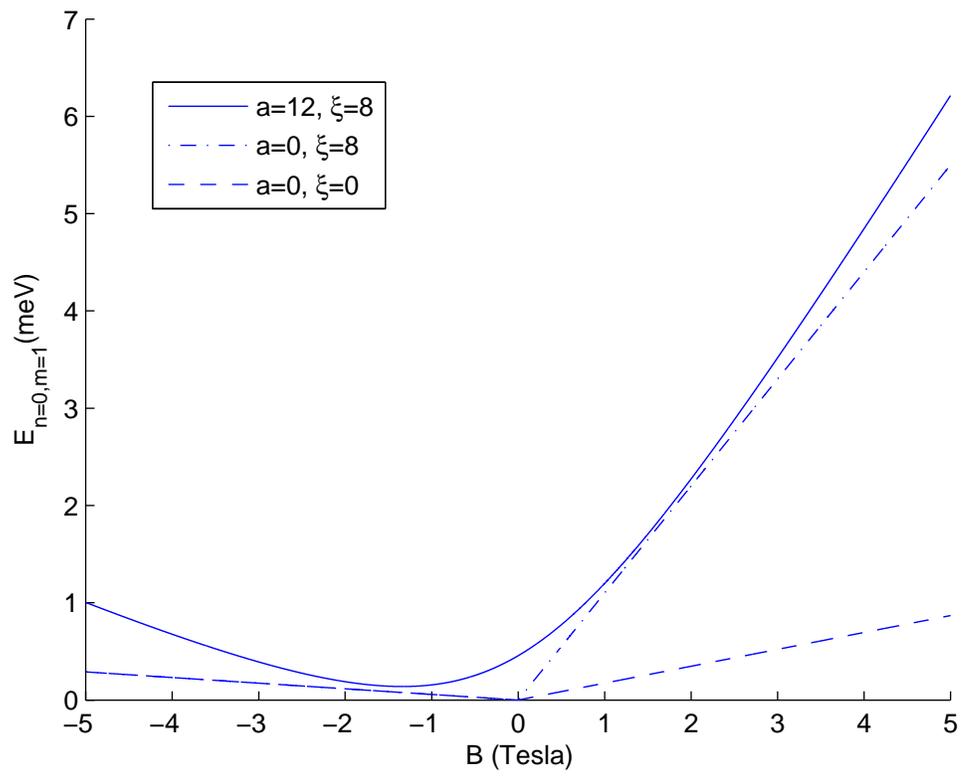
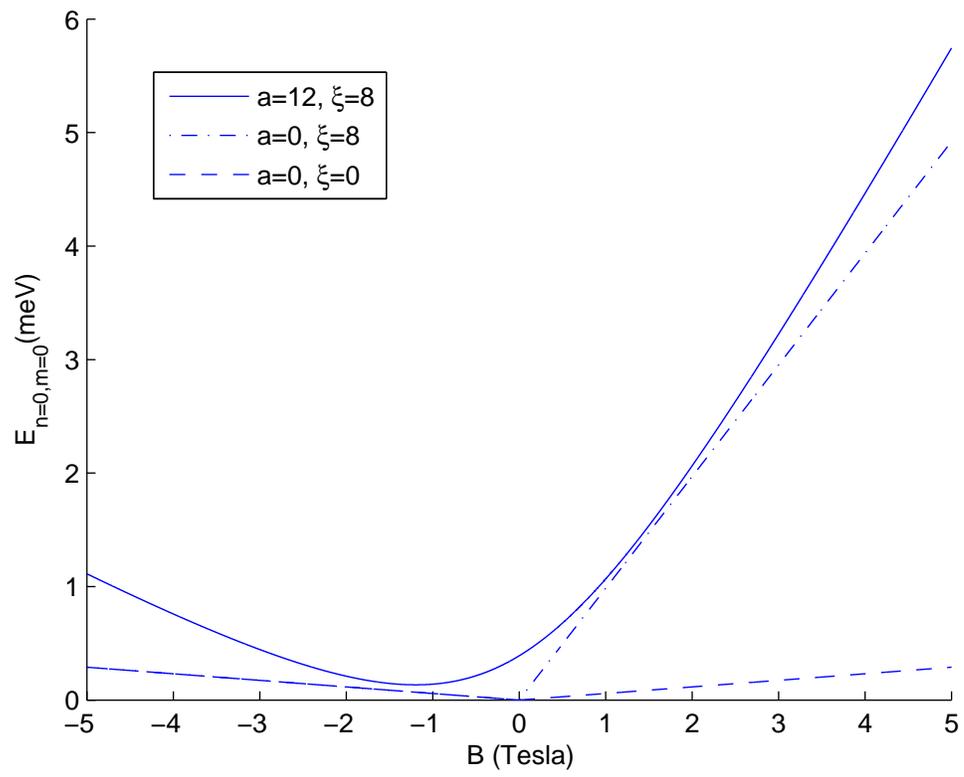
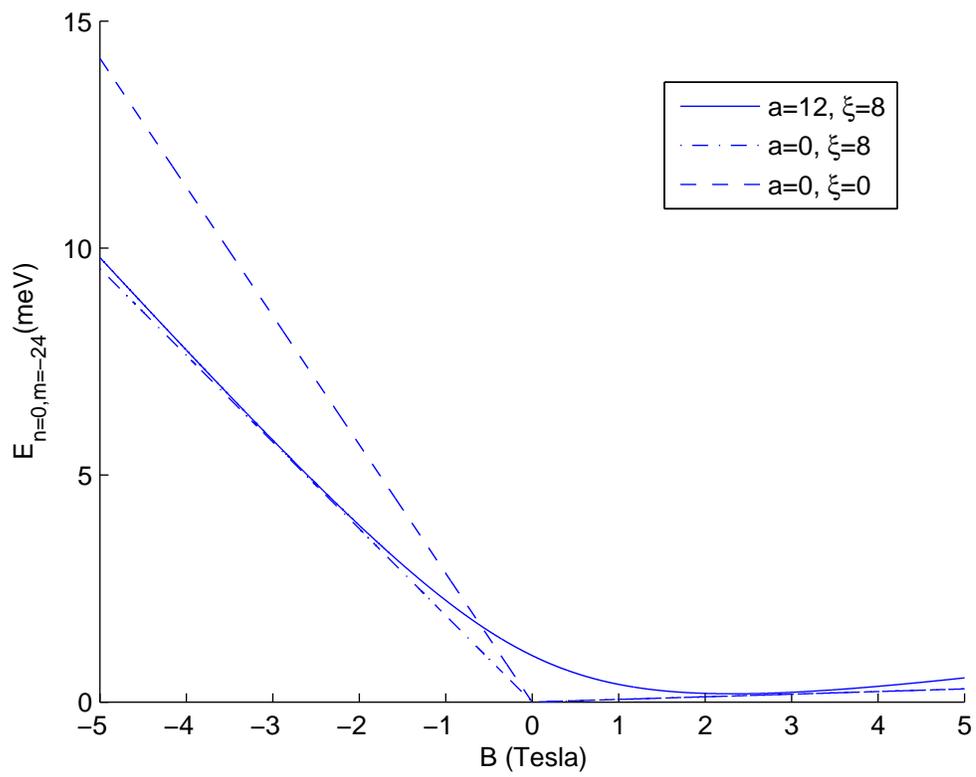
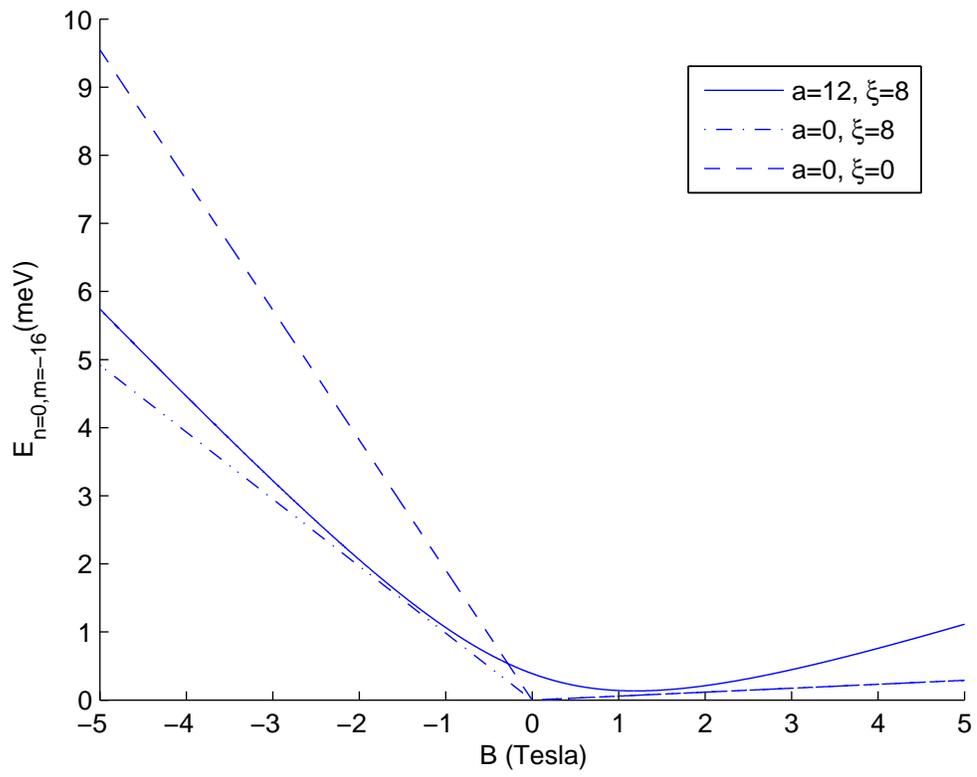
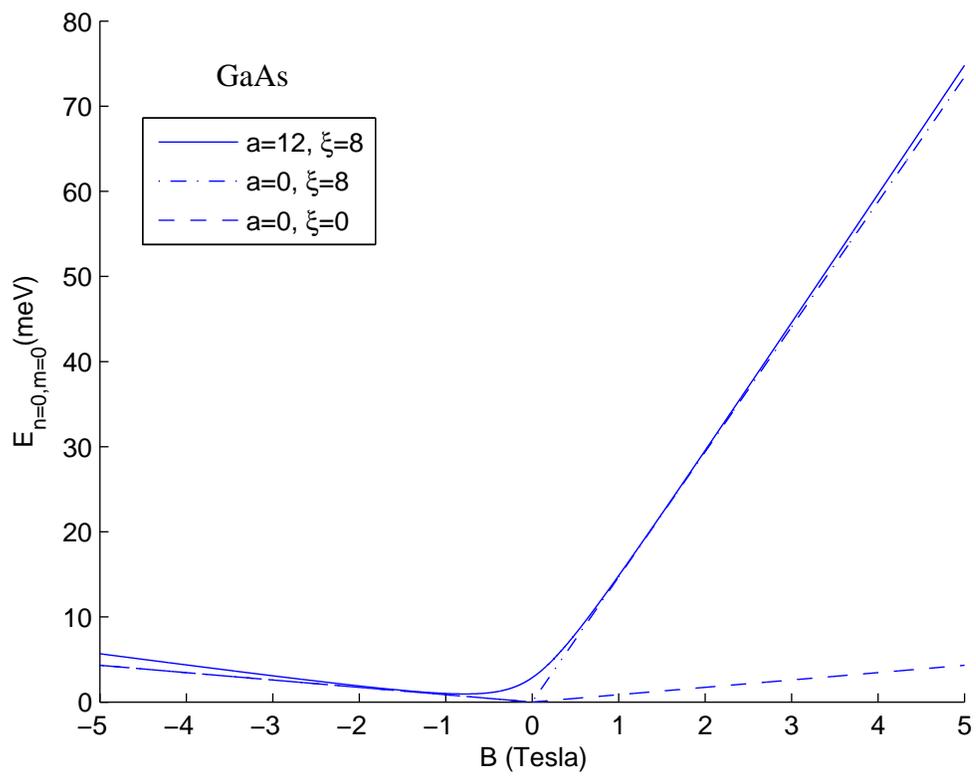
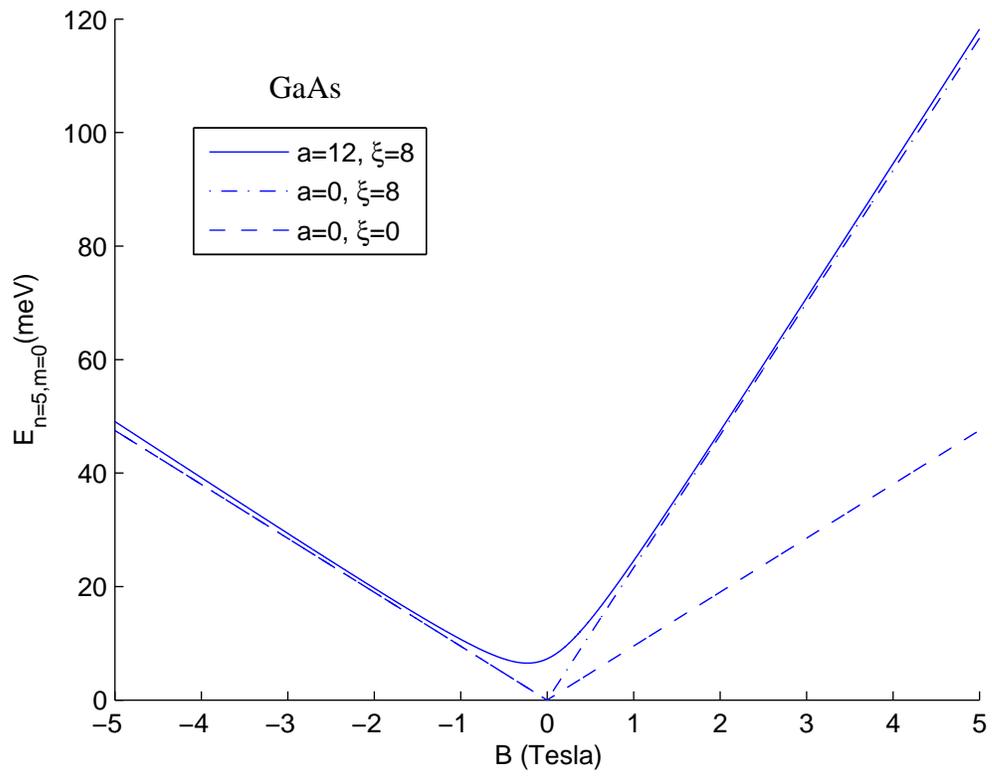
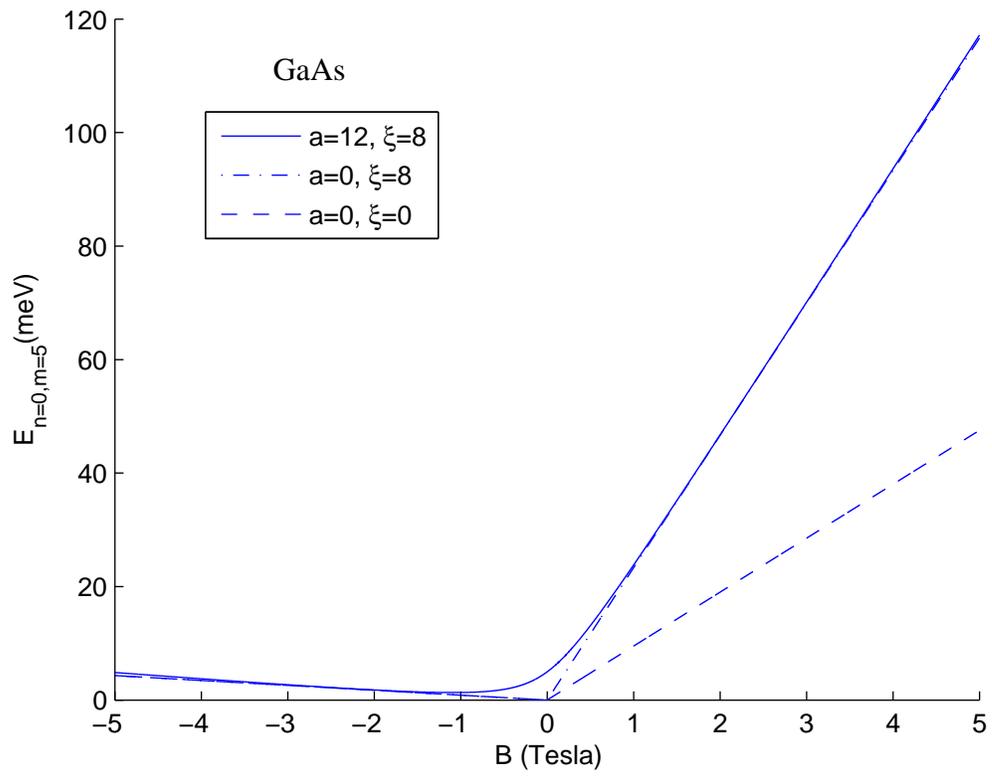
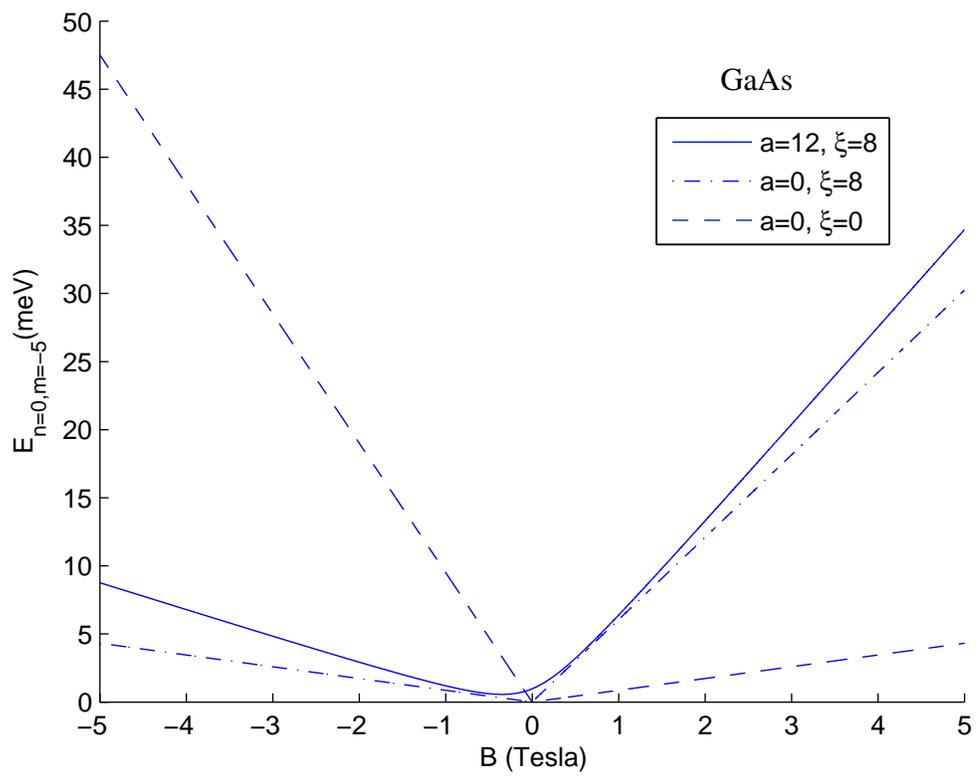
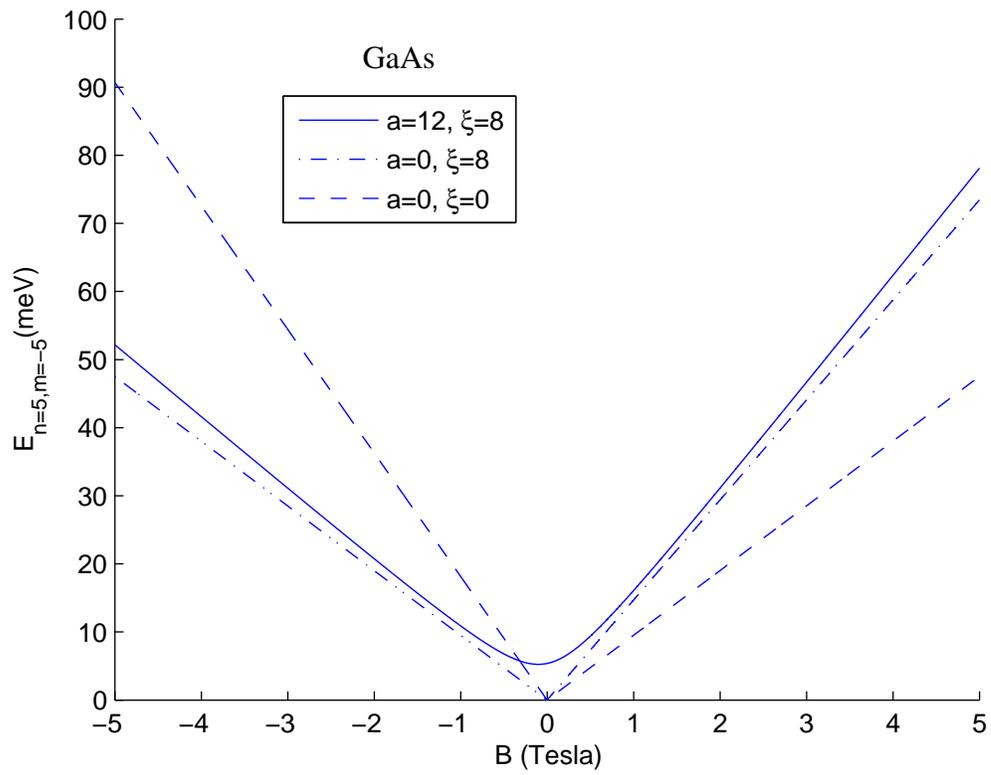
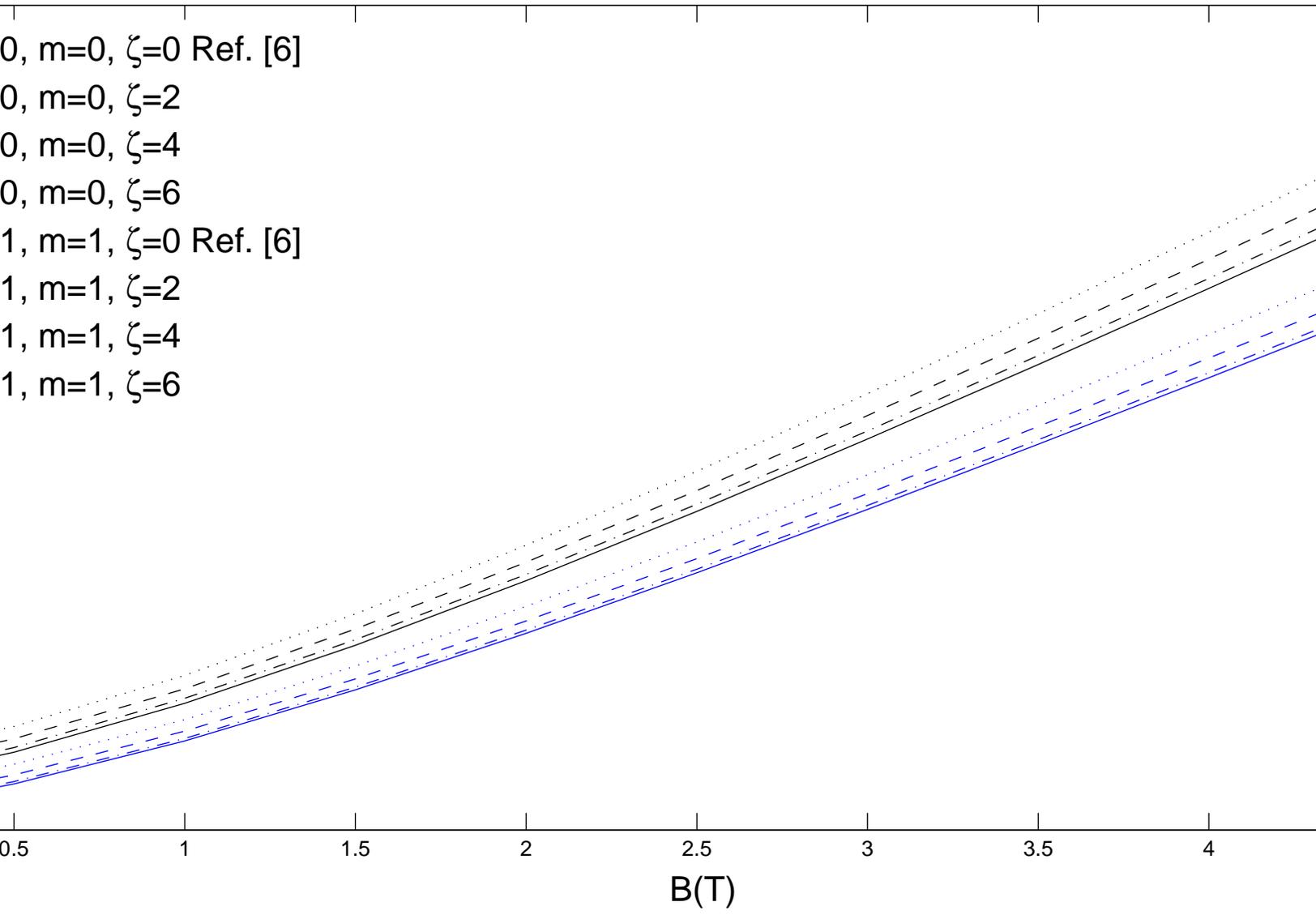
\end{document}